# A Van der Waals Moiré Bilayer Photonic Crystal Cavity

Lesley Spencer[1,2,†], Nathan Coste[1,2,†,*], Xueqi Ni[3], Seungmin Park[1,4], Otto C. Schaeper[1,2], Young Duck Kim[4], Takashi Taniguchi[5], Kenji Watanabe[6], Milos Toth[1,2], Anastasiia Zalogina[1,2], Haoning Tang[7,8] and Igor Aharonovich[1,2*]

[1] School of Mathematical and Physical Sciences, University of Technology Sydney, Ultimo, New South Wales 2007, Australia

[2] ARC Centre of Excellence for Transformative Meta-Optical Systems, University of Technology Sydney, Ultimo, New South Wales 2007, Australia

[3] Department of Physics, National University of Singapore 119077 Singapore

[4] Department of Physics, Kyung Hee University, Seoul 02447, Republic of Korea

[5] International Center for Materials Nanoarchitectonics, National Institute for Materials Science, 1-1 Namiki, Tsukuba, 305-0044, Japan

[6] Research Center for Functional Materials, National Institute for Materials Science, 1-1 Namiki, Tsukuba, 305-0044, Japan

[7] Department of Electrical Engineering and Computer Science, University of California at Berkeley, Berkeley, CA 94720, USA

[8] School of Engineering and Applied Sciences, Harvard University, Cambridge, MA 02138, USA

* Correspondence to: nathan.coste@uts.edu.au, igor.aharonovich@uts.edu.au

**Abstract**

*Enhancing light-matter interactions with photonic structures is critical in classical and quantum nanophotonics. Recently, Moiré twisted bilayer optical materials have been proposed as a promising means towards a tunable and controllable platform for nanophotonic devices, with proof of principle realisations in the near infrared spectral range. However, the realisation of Moiré photonic crystal (PhC) cavities has been challenging, due to a lack of advanced nanofabrication techniques and availability of standalone transparent membranes. Here, we leverage the properties of the van der Waals material hexagonal Boron Nitride to realize Moiré bilayer PhC cavities. We design and fabricate a range of devices with controllable twist angles, with flatband modes in the visible spectral range (~ 450 nm). Optical characterization confirms the presence of spatially periodic cavity modes originating from the engineered dispersion relation (flatband). Our findings present a major step towards harnessing a two-dimensional van der Waals material for the next-generation of on chip, twisted nanophotonic systems.*

Confining light in photonic structures enables the engineering of light-matter interactions at the nanoscale, leading to fascinating applications in classical and quantum technologies[1-3]. Over the last few decades, optical resonators, metasurfaces based on Bound-States in the Continuum, and photonic crystal (PhC) cavities have been at the forefront of nanoscale light manipulation[4-8]. They enabled miniaturised light emitting devices, nanoscale lasers[9], and have been instrumental in realising on-chip quantum photonic architectures used to enhance and manipulate quantum light sources[10].

A recent pivot in nanophotonics has been motivated by the observation of Moiré superconductivity and correlated phenomena in twisted bilayer graphene[11, 12]. In photonic media, Moiré engineering via the twist degree of freedom can enable nontrivial topological properties and new approaches to light confinement[13-19]. Indeed, theoretical proposals have been put forward to control material dispersion relations, and to realise flatband engineering[20-24]. To this end, several key implementations have been demonstrated, including the realization of nanoscale lasers[25-28], Moiré metasurfaces for beam steering[29] via dispersion control, polariton propagation in twisted bilayers[30] and active tuning of spontaneous parametric down-conversion[31].

However, the main challenge impeding progress of twist-optics stems from difficulties in transferring and aligning patterned nanoscale optical slabs. Critically, these underpin the realisation of Moiré PhC cavities that can achieve a maximal light confinement in a small modal volume[17, 31, 32]. These challenges have been, to date, overcome by patterning the twisted design into a single slab of material[25, 32, 33]. Whilst this approach works, it is permanently static and lacks the ability to tune and reconfigure a twisted multi-slab system, which makes twist-optics so attractive[34].

Here, we design, fabricate and characterise true Moiré PhC cavities by patterning and assembling two slabs of hexagonal boron nitride (hBN). hBN offers an outstanding opportunity for advanced twist optics, being a transparent Van der Waals[35] material with a bandgap of 6 eV. Hence, hBN is an appealing platform for fabricating Moiré PhC cavities from standalone hBN PhC slabs. Utilising advanced nanofabrication techniques, we pattern individual PhC slabs and assemble them into twisted, bilayer Moiré PhC cavities. Our approach illustrates the capabilities of Van der Waals materials to engineer innovative devices towards advanced nanophotonic technologies.

Figure 1 illustrates the concept of our devices, fabricated from two hBN PhCs. The inset of Fig 1a shows the crystallographic lattice of hBN. When two identical PhC slabs are stacked and twisted, a periodic superlattice and a well-defined Moiré pattern forms at specific commensurate twist angles. The resulting pattern consists of AA, AB, and BA-stacked regions with macroscopic periodicities controlled by the twist angle, as illustrated in Fig. 1b. The periodicity of the Moiré superlattice $b$ is defined by the lattice constant of the single PhC $a$ and the relative twist angle $\theta$ between the two slabs: $b = a/[2sin\left(\frac{\theta}{2}\right)]$. The superlattice periodicity leads to a modified dispersion relation that can result in a flatband[36, 37], featuring zero group velocity (light trapping) and light confinement at the AA sites. Unlike conventional PhC cavities, which confine light through defect engineering, bilayer PhC cavities enable optical confinement via momentum-free trapping of Bloch waves, effectively increasing the photonic density of states[38]. As the twist angle $\theta$ decreases, mode localization in the AB and BA regions shifts toward the AA regions, leading to stronger light confinement, as schematically illustrated in Fig. 1c.

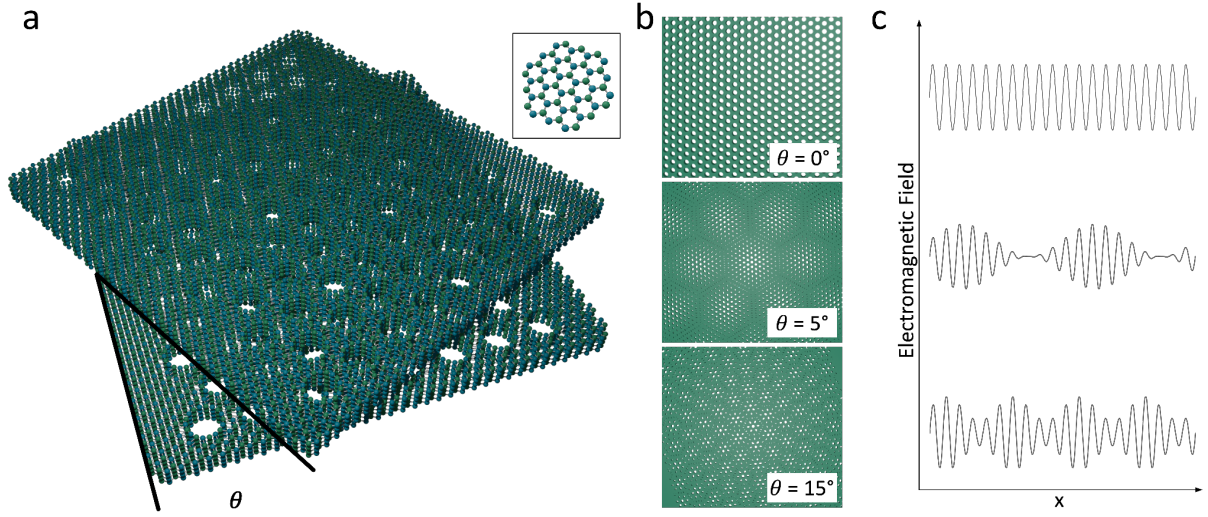

*Figure 1: Concept illustration of twisted bilayer hBN PhC cavity slabs. a, Schematic showing a stack of two PhCs with twist angle θ. Each PhC is an identical array of air holes in a slab of the van der Waals material hBN. Crystallographic structure of an hBN monolayer is shown schematically as an inset. b, Schematics of the twisted PhC slabs and the resulting Moiré pattern at angles of 0°, 5° and 15° degrees, displaying different periodicities at θ=5° and θ=15°. c, Amplitude of the electromagnetic fields modulated by the corresponding twisted PhCs in (b).*

The band structures from bilayer systems are based on interlayer and intralayer coupling mechanisms and described by an effective four-component Hamiltonian and a low-energy tight-binding model which describes minibands of the Moiré superlattice. The Hamiltonian, considering only nearest coupling for a single layer two-dimensional hexagonal lattice, can be described as:

$$H(k) = t \sum_i e^{ik \cdot r_i} \quad (1)$$

where $k$ is the wave vector, $t$ is the coupling strength, $r_i$ is the nearest neighbour vectors with $r_1=a(1,0)$, $r_2=a(1/2,\sqrt{3}/2)$, and $r_3=a(-1/2,\sqrt{3}/2)$, and $a$ is the lattice constant. For bilayer structures, the total Hamiltonian becomes $H = [H_1(k), V; V^*, H_2(k)]^{39}$, where $H_{1(2)}(k)$ represents a single layer Hamiltonian for the first (second) layer, and $V$ is the interlayer coupling strength. The interlayer coupling strength can be tuned via the distance between two PhC slabs. We simulated a few such distances between two PhC layers and found the interlayer distance h = 50 nm gives rise to a flat dispersion within a photonic bandgap (see Fig. S(1-3) in Supplementary Information).

We begin by designing the band structures of the hBN PhCs. The bands were simulated numerically using three-dimensional finite-element methods (COMSOL Multiphysics). The twisted bilayers consist of PhC layers with refractive index of *n=1.8* (matching the hBN refractive index) which were scaled for a wavelength of ~ 450 nm. The simulated design had a

$C_{6v}$ symmetry-protected triangular lattice with a constant $a = 270$ nm, air hole diameter $d = 162$ nm, and a PhC slab thickness of $h = 135$ nm. The chosen target wavelength of 450 nm stems from the fact that hBN hosts quantum emitters in this spectral range which can be engineered deterministically[40]. Hence, there is a clear need for cavity engineering with resonances at this wavelength.

The dispersion relations were calculated for various twist angles $\theta$ (see Fig. S2 in Supplementary Information). Fig. 2(a, b) show the Purcell factor obtained for twist angles of 13° and 6°, respectively, across the full flatband area for a range of 650 - 660 THz (454 - 461 nm).

The interlayer tunnelling mechanism, investigated via slab spacing induces coupling between the two layers which results in a guided resonance (see Fig. 2c). The TE mode illustrated in Fig. 2c propagates through the air holes in the bottom layer and weakly couples to the air holes in the upper layer, mimicking the electron hopping between atoms in a vdW material.

The vertical and lateral confinement provides twist-angle tunable, momentum-free trapping of Bloch waves. Our simulations show that the flat-band modes have higher Q-factors ($>10^5$) at smaller angles and radiate only weakly in the vertical direction, despite lying above the light cone. Fig 2d shows the frequency at the $\Gamma$ point as a function of the twist angle. Note that $Q$ remains high regardless of the twist angle in the immediate proximity of the $\Gamma$ point .

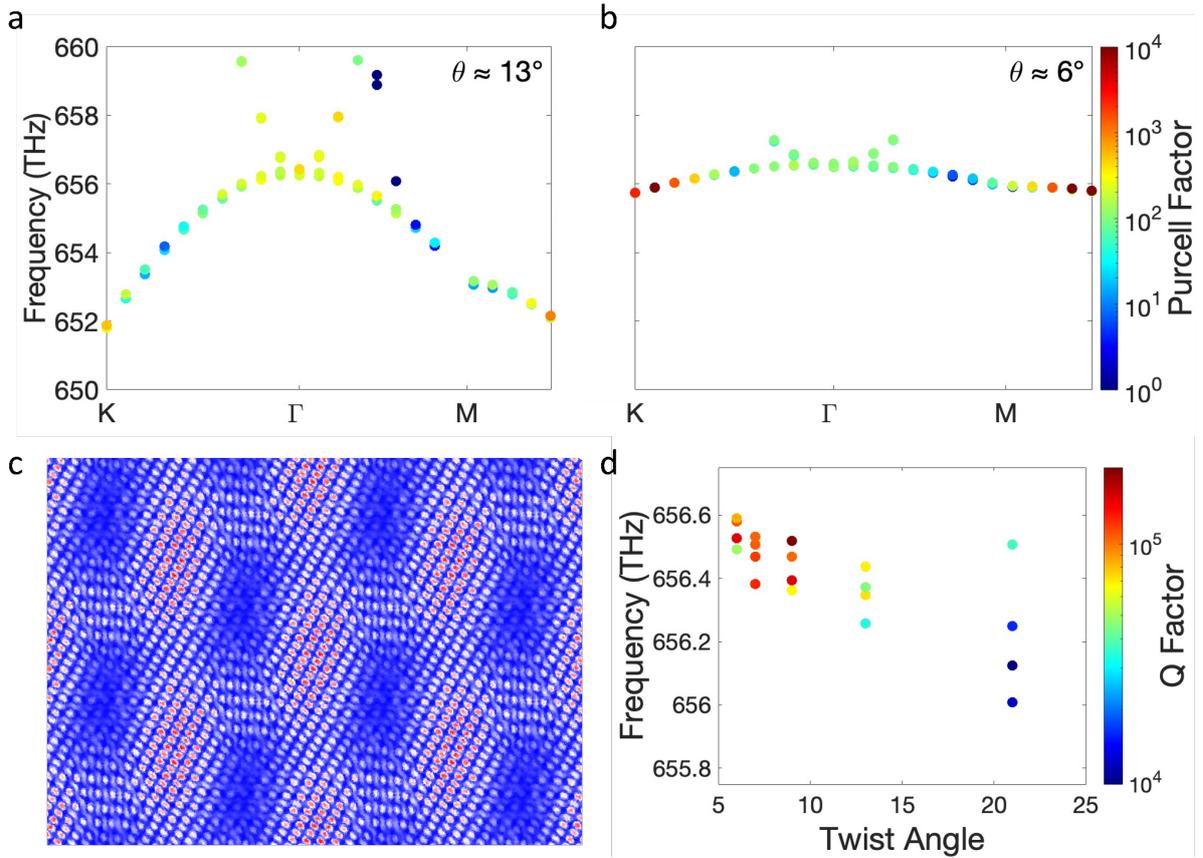

*Figure 2: Numerical simulations of twisted bilayer PhC slabs.* **a,** Band structure diagrams for TM modes as a function of wave-vector **k** and frequency for a twist angle of $\theta=13°$ and **b,** $\theta=6°$; The color bar represents the corresponding Purcell factor. **c,** Electromagnetic distribution of the mode profile showing light confinement at AA-stacked regions. **d,** Q factor dependency on frequency and twist angles at the $\Gamma$ point.

To fabricate the individual PhC slabs, hBN flakes were mechanically exfoliated onto an silicon dioxide substrate (Fig. 3a(i)), and a few flakes were selected based on the desired thickness range of 120 - 200 nm (Fig. 3a (ii)). The PhC patterns were then defined by electron beam lithography and fabricated in the hBN flakes using reactive ion etching (Fig. 3a (iii)). To assemble the twisted bilayer PhC cavities, two PhC slabs are stacked one on top of another at different angles using a homebuilt aligned transfer setup (see Fig. S4 in supplementary information for details). Using this setup, the upper PhC slab was picked up from its substrate by a polydimethylsiloxane stamp (PDMS) coated in polycarbonate[41], as shown in Fig. 3a (iv). It was then aligned with the lower PhC slab at a chosen angle (Fig 3a (v)) and deposited to create the final bilayer. The Moiré superlattice formed in the twisted PhC slabs is schematically shown in Fig 3a (vi).

Fig. 3b shows an optical image of the patterned flake (yellow) selected to be the 'Upper PhC slab' on an $SiO_2$ substrate. Fig. 3c shows the flake selected to be the 'Lower PhC slab'. The outline in Fig. 3c indicates the patterned of hBN). Fig 3d shows both hBN slabs part way through the aligned transfer process. This optical image was taken through the stamp, as illustrated in Fig 3a (iv). It shows the upper PhC slab attached to the stamp and aligned to a twist angle of 15°, above the lower PhC slab. Fig 3e shows the final stack after the polymer removal with both upper and lower PhC slabs outlined for clarity.

A scanning electron microscope (SEM) image of a single fabricated PhC slab is shown in Fig. 3f. Fabrication and transfer of patterned PhCs required the development of an entirely new chemical dry etch process to be developed, as discussed in Supplementary Section 2. This process enabled the structures to be transferred without chemical removal of the substrate that is often required[42].

Based on our design, the hBN PhC has a lattice constant $a$ of 270 nm (Fig. 3f). An SEM image of the two stacked PhC slabs forming the Moiré lattice with a twist angle of $\theta=21°$ is shown in Fig. 3g. The image clearly demonstrates a honeycomb-like Moiré pattern with a periodicity (shortest distance between AA sites) of ~ 720 nm, nearly twice the single PhC lattice constant $a = 270$ nm. This aligns well with the calculated periodicity of 740 nm. As the twist angle $\theta$ decreases, the Moiré periodicities increase in size. The effect of the twist angle on the size of the Moiré periodicity is visible in Fig. S5 & S7 in Supplementary Information.

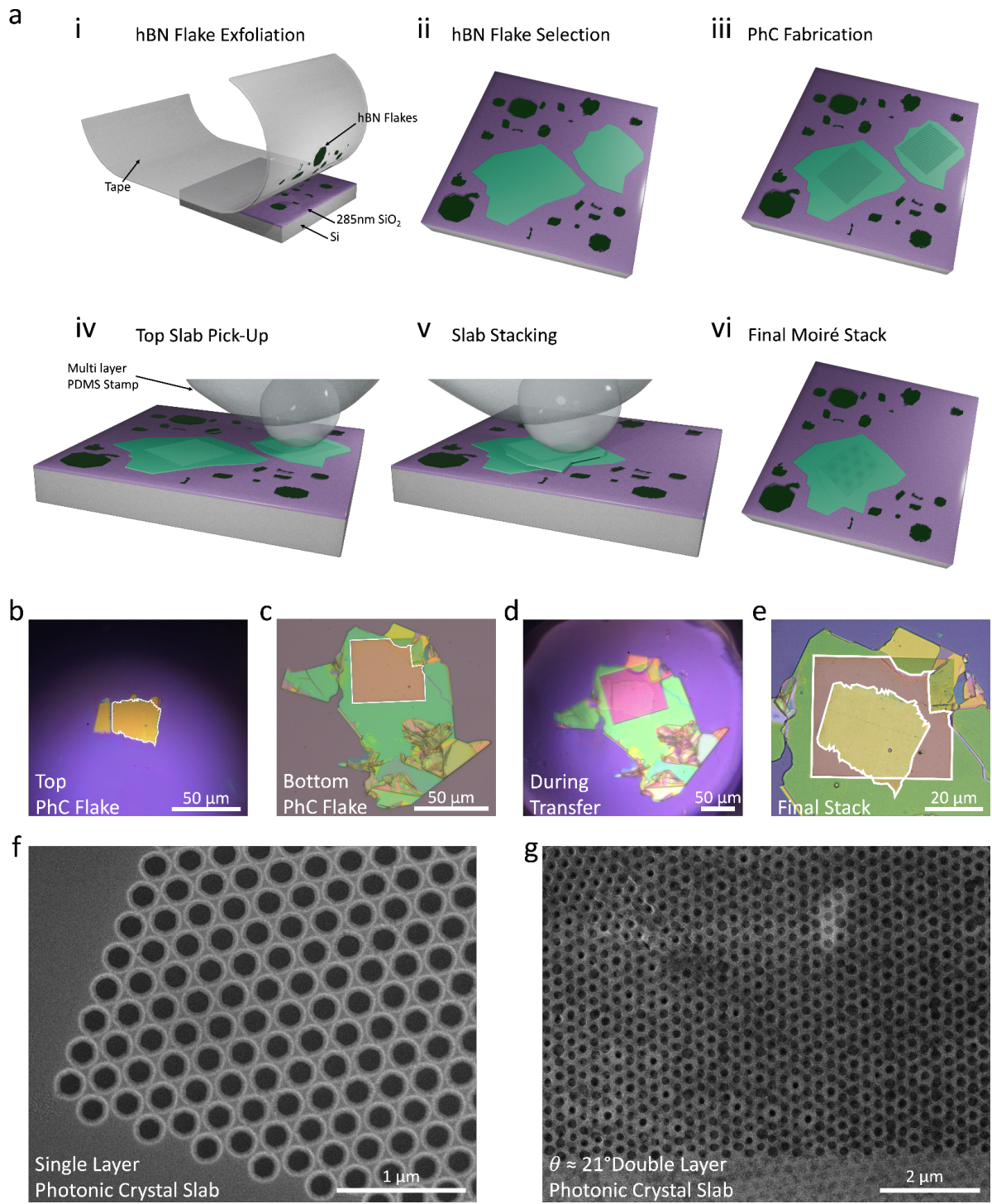

***Figure 3: Fabrication of single and twisted hBN Moiré PhC cavities. a,*** *Schematic of the fabrication steps of twisted hBN Moiré bilayer photonic crystal (PhC) cavities.* ***i****, Exfoliation of hBN flakes.* ***ii****, Flake selection.* ***iii****, Design patterning of a PhC.* ***iv****, Pick-up the patterned flake with polydimethylsiloxane (PDMS) stamp.* ***v****, Flake twist and transfer onto the second patterned flake.* ***vi****, Final Moiré PhC cavity.* ***(b-e)*** *Optical images of the PhC slabs (outlined) during the transfer process.* ***(b)*** *hBN flake with PhC pattern (yellow highlighted) selected to be the upper slab.* ***(c)*** *hBN flake with PhC pattern (brown and outlined) serving as the lower slab.* ***(d)*** *image through the transfer system and polymer stamp showing the alignment of the top*

PhC slab (transparent & attached to the stamp) with the bottom PhC slab on the substrate. (**e**) The final twisted Moiré bilayer PhC after polymer removal with both upper and lower PhCs outlined in white for clarity. (**f**) A scanning electron microscope (SEM) image of a single PhC slab and (**g**) SEM a image of a twisted Moiré bilayer (two stacked and twisted PhC slabs) at an angle of $\theta=21°$, with the bottom slab visible in the top left corner.

Finally, we optically characterize the fabricated Moiré bilayer PhCs with twist angles of $\theta=15°$ and $\theta=5°$, respectively. Figure 4a shows the experimental setup. We perform resonant scattering measurements in a cross-polarization configuration: the excitation laser, a spectrally broad pulse ranging from 440 nm to 500 nm generated by a supercontinuum laser, is sent through a polarizing beamsplitter and a set of waveplates to excite the flatband resonant mode with linear polarization[43]. The part of the beam that couples to the flatband resonance undergoes polarization rotation due to the material birefringence, and is then reflected by the polarizing beamsplitter towards the single mode fiber collection. In contrast, light that scatters at the top of the device and does not couple to the cavity is reflected back towards the laser path, hence filtered out from the collection, ensuring that only light coupled with the flatband mode is collected.

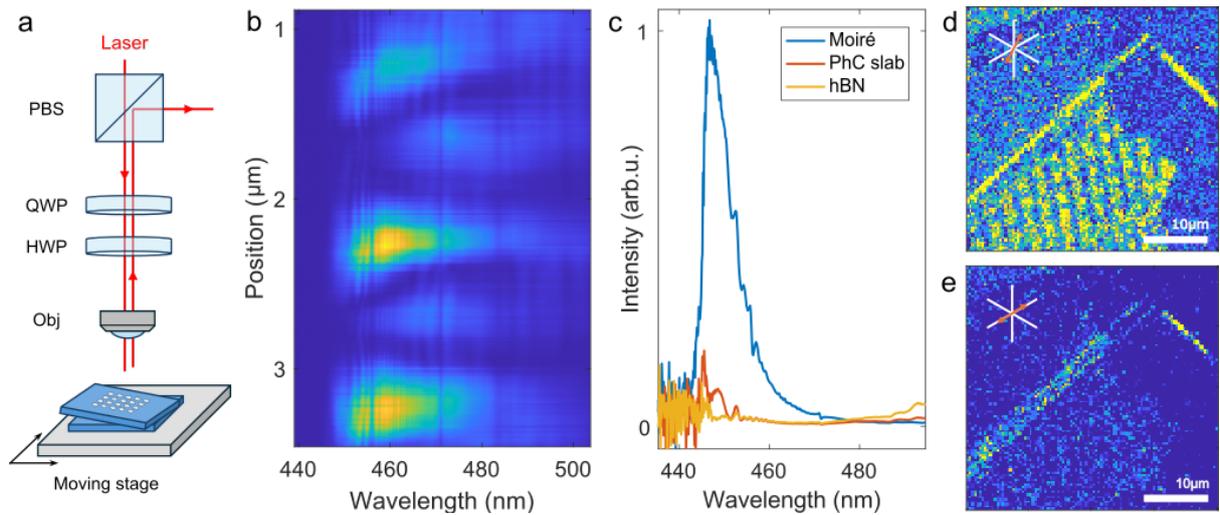

*Figure 4: Optical measurements of twisted Moiré bilayer PhCs.* **a**, Sketch of the experimental setup used to characterize the devices (PBS: polarizing beamsplitter, QWP: quarter waveplate, HWP: half waveplate, Obj: microscope objective). **b**, Optical measurements for the twisted bilayer PhCs with twist angles of 15°. Cross-polarized spectra are shown, while scanning the laser beam across the sample in one direction. Strong emission is corresponding to the periodicity of the Moiré electromagnetic field for this twist angle. **c**, Normalized cross-polarized spectra showing the flatband resonant mode (blue) for the twisted Moiré bilayer PhC (15°). As a c comparison, spectra were taken from a single PhC slab (red), and unpatterned hBN flake (yellow). **d, e**, Optical measurements for the twisted bilayer PhCs with twist angles of 5°. 2D intensity maps recorded from the cross-polarized signal on an avalanche photodiode,

*for two excitation polarizations separated by 30 degrees (see text for details). Maxima and minima are clearly visible in (d) and (e), respectively. The inset shows a sketch of the cavity axes (white lines) and excitation polarization (orange arrow).*

To spatially map the flatband modes of the twisted Moiré bilayer PhCs, we record the spectra of the cross-polarized signal as a function of the laser beam position on the sample. To do so, we scan the sample in one direction below the laser spot and record spectra every 100 nm over several microns across the sample. The resulting spatially resolved spectral map is shown in Fig. 4b for the *15°* Moiré bilayer PhC device. We observe a cross-polarized signal with a 1-µm spatial periodicity, which is consistent with the expected Moiré periodicity lattice constant for this twist angle. This result convincingly demonstrates the localisation of the electromagnetic field at the cavity maxima, corresponding to the twisted Moiré PhC cavity.

Figure 4c presents the spectrum extracted from Fig. 4b at a resonance position and normalized by the incident laser spectrum, revealing the characteristic flatband mode at 447 nm. For reference, the same measurements are taken on the PhC slab and unpatterned hBN, evidencing no such mode. The flatband resonance shows an asymmetric lineshape that can be explained by the parabolic dispersion relation (Fig. 2a) and geometry of the measurement: given the laser path is perpendicular to the sample, light couples more strongly to the propagating modes close to the **Γ** point, which results in a stronger contribution of these modes at higher frequency in the measured spectra.

The full width at half maximum (FWHM) of ~ 6 nm of the cavity mode corresponds to a $Q$ factor of 70. While low compared to traditional PCCs, this is the first measurement of any $Q$ value from a twisted Moiré PhC cavity in the visible range. The measured $Q$ factor matches the modelling well, and is relatively low due to the absence of an air gap between the two slabs. The ideal device should comprise an air gap of ~ 50 nm between the slabs, which is expected to increase the $Q$ factor by several orders of magnitude.

Further, we investigate the polarization dependency of the flatband modes. This is important as the modes should correspond to the symmetry of the triangular lattice. We turn to the bilayer PhC device with a twist angle of 5°. Here, we set the laser frequency at the flatband resonance and measure a 2D map of the cross-polarization signal intensity on an avalanche photodiode. We perform this measurement for two different polarization angles separated by 30 degrees, corresponding to a maximum (Fig 4d) and minimum (Fig 4e) of the cross-polarized signal intensity. This indicates that the flatband modes are linearly polarized 60 degrees apart, consistent with the symmetry of the triangular lattice of the PhC slabs.

To summarize, we have demonstrated the design, fabrication and measurement of twisted hBN Moiré bilayer PhCs in the visible spectral range. The van der Waals nature of hBN is an enabler for the implementation of twisted, patterned photonic structures, due to its controllable thickness, and a practical dry transfer with a controllable twist angle. Such cavities are expected to show promising figures of merits (high Q factor and small mode volume), which could be obtained in our devices by introducing an air gap between the slabs.

The design of such devices offers a flexible choice of materials and lattice parameters, to engineer the dispersion relation on-demand and achieve flatband resonances at tunable wavelength and bandwidth. Going forward, the integration of quantum emitters with hBN twisted Moiré PhCs could lead to exciting phenomena including strong coupling and ultra high Purcell enhancement to study new cavity quantum electrodynamic regimes[44, 45]. Furthermore, extending these designs to other materials such as Transition Metal Di-Chalcogenides (TMDCs) that possess higher refractive indexes may offer even stronger light localisation, and enable low threshold lasing and fundamental studies of many body systems[45, 46]. Combination of strain tuning (using piezo elements or stretchable materials) may provide additional levers to achieve a desired interplay between light and matter, and be particularly useful for tunability and reconfigurability.

## Methods

**Numerical Simulations.** Band structures of PhCs were designed and numerically simulated using three-dimensional finite-element methods (COMSOL Multiphysics). PhC slabs as a square lattice with circular air holes of diameter $d = 162$ nm and a triangular lattice constant $a = 270$ nm are modelled with a refractive index $n = 1.8$, which is representative of boron nitride.

In the COMSOL simulation, we consider two adjacent PhC slabs twisted by an angle relative to one another. This produces a Moiré superlattice with a macroscopic periodicity of distinct AA and AB/BA stacking regions that grow in size as the angle decreases. Because the COMSOL finite-element calculation relies on the existence of Bloch waves, we ensure that the structures created by twisting two lattices relative to each other are exactly commensurate by considering only specific twist angles:

$$\theta = 2\arcsin\left(\frac{1}{2\sqrt{3n^2 + 3n + 1}}\right)$$

We used the periodic boundary conditions for the boundaries of the Moiré superlattice and the perfect matching layers for the out-of-plane radiative directions. Then we plot the frequency of the quasi-TM eigenmodes into the band structure. The imaginary part of the eigenfrequency indicates the Q-factor of the eigenmode.

**Fabrication.** To fabricate the designed PhC structures, 140-nm thick hBN flakes were mechanically exfoliated with 3M Scotch tape on $SiO_2$/Si substrates. Pristine hBN flakes were sourced from the National Institute for Materials Science (NIMS). A selection of hBN flakes thicknesses was based on a combination of optical contrast and atomic force microscope measurements. A layer of electron beam resist (AR-P 6200 series, CSAR62, AllResist GMBH) was spin-coated onto the exfoliated hBN flakes at 4000 rpm for 60 s, followed by baking on a hotplate at 180 °C for 3 min. The resist was then patterned using electron beam lithography (Elionix ELS-F125), with an area of dose of electron beam exposure of 600 µC/cm², at 125 kV and 1 nA. The exposed pattern was developed in AR 600-546 for 40 s, and rinsed in IPA for 20 s. The electron beam resist served as a mask for transferring the designed pattern to the material using a reactive ion etching in the ICP-RIE system (Trion) at 20 mT, 40 sccm $SF_6$, 5 sccm Ar, 40 W RF, and 400 W ICP with the etch rate of $\approx$ 7.5 nm/s. The RF power was reduced to prevent 'welding' of the flake to the substrate (see Section 2 in Supplementary Information for more details). Finally, the electron beam resist mask was removed with oxygen plasma in the ICP-RIE system (Trion) at 100 mT, 100 sccm $O_2$ 20 W RF, and 500 W ICP for 10 s, followed by soaking in an 80 °C Remover AR 600-71 for 2 min. Bilayer hBN PhC slabs were assembled and twisted using a dry transfer method with a custom-built setup and a polydimethylsiloxane (PDMS; SYLGARD™ 184 Silicone Elastomer, Dow, MI) stamp coated with a thin film of polycarbonate (PC) polymer (see Figure S4 in Supplementary Information). The stamp was used to pick up one of the PhCs patterned into hBN flakes at 80-140°C, align it at angle $\theta$ with the lower patterned hBN PhCs, and release it at 200-240 °C. Residual PC polymer was removed by dissolving it in chloroform overnight. The final structures were inspected, and scanning electron microscope images were obtained using a Thermo Fisher Scientific Helios G4 SEM.

**Optical Characterization.** Samples were characterised using a lab built, resonant, confocal setup to measure the resonant modes from the Moiré PhC in a cross-polarization configuration. The light source is a supercontinuum laser NKT Fianium FIU-15 with tunable VARIA frequency filter. The light is focused onto the sample through a 100× 0.9 NA objective. The resulting signal is collected either via a single mode fiber or detected by a free-space avalanche photodiode.

The incident laser polarization is controlled using a polarizing beamsplitter (PBS) and a set of half and quarter waveplates. The incident light with linear polarization interacts with the Moiré PhC, where only the resonantly coupled modes experience a polarization rotation due to birefringence of the material. The configuration of waveplates and PBS ensures that only light that has undergone polarization rotation is reflected by the PBS towards the setup collection. The angle of the linear polarization for the excitation is chosen to be halfway between the cavity axes which maximizes the collected signal.

The samples are mounted on a NanoCube XYZ positioner. We scan the sample below the laser spot and record spectra or intensity as a function of position to spatially map the cavity resonances.

**Acknowledgments**

The authors acknowledge financial support from the Australian Research Council (CE200100010, FT220100053, DP240103127) and the Office of Naval Research Global (N62909-22-1-2028). The authors acknowledge Takashi Taniguchi and Kenji Watanabe (the National Institute for Materials Science) for providing the hBN crystals. S.P and Y.D.K was supported by the National Research Foundation of Korea (NRF) grant funded by the Korea government (MSIT) (2022M3H4A1A04096396, RS-2023-00254055). K.W. and T.T.

acknowledge support from the JSPS KAKENHI (Grant Numbers 21H05233 and 23H02052) and World Premier International Research Center Initiative (WPI), MEXT, Japan. The authors acknowledge the use of the fabrication facilities as well as scientific and technical assistance from the Research and Prototype Foundry Core Research Facility at the University of Sydney, being a part of the NCRIS-enabled Australian National Fabrication Facility (ANFF), and the UTS facilities, being a part of the ANFF-NSW node.